%%%%%%%%%%%%%%%%%%%%%%%%%%%%%%%%%%%%%%%%%%%%%%%%%%%%%%%%%%%%%%%%%%%%%%
%%%%%%%%%%%%%%%%%%%%%%%%%%%%%%%%%%%%%%%%%%%%%%%%%%%%%%%%%%%%%%%%%%%%%%
%%%                                                                %%%
%%%         .         .           ..           .         .         %%%
%%%       Single query learning from abelian and non-abelian       %%%
%%%                    Hamming distance oracles                    %%%
%%%                                                                %%%
%%%              David A. Meyer and James Pommersheim              %%%
%%%                                                                %%%
%%%                                                                %%%
%%%%%%%%%%%%%%%%%%%%%%%%%%%%%%%%%%%%%%%%%%%%%%%%%%%%%%%%%%%%%%%%%%%%%%
%%%%%%%%%%%%%%%%%%%%%%%%%%%%%%%%%%%%%%%%%%%%%%%%%%%%%%%%%%%%%%%%%%%%%%
%%%                                                                %%%
%%%                                                                %%%
%%%                       Typesetting notes                        %%%
%%%                                                                %%%
%%%                                                                %%%
%%% Plain TeX with a few simple macros at the beginning.           %%%
%%%                                                                %%%
%%% If AMS fonts are not loaded, comment the next line and         %%%
%%%  uncomment the following one.                                  %%%
\font\bbb=msbm10 \font\bbs=msbm7                                   %%%
%\def\bbb{\bf} \def\bbs{\bf}                                       %%%
%%%                                                                %%%
%%%                                                                %%%
%%%%%%%%%%%%%%%%%%%%%%%%%%%%%%%%%%%%%%%%%%%%%%%%%%%%%%%%%%%%%%%%%%%%%%
%%%%%%%%%%%%%%%%%%%%%%%%%%%%%%%%%%%%%%%%%%%%%%%%%%%%%%%%%%%%%%%%%%%%%%

\overfullrule=0pt

%\input epsf.tex
%\input pstricks  
%\input color
%\definecolor{Green}{rgb}{0.25,0.75,0.25}
%\definecolor{review}{rgb}{0.25,0.75,0.25}
%\definecolor{exercise}{rgb}{1,0,0}
%\definecolor{lightgray}{rgb}{0.5,0.5,0.5}

\def\C{\hbox{\bbb C}} \def\sC{\hbox{\bbs C}}

\def\Z{\hbox{\bbb Z}} \def\sZ{\hbox{\bbs Z}} 
\def\Tr{{\rm Tr}}

\font\tinyss=cmss6
\font\ss=cmss10

\font\ssbf=cmssbx10

\def\IC{{\sl Inform.\ Control}}

\def\IRETIT{{\sl IRE Trans.\ Inform.\ Theory}}

\def\ML{{\sl Machine Learning}}

\def\PIEEEL{{\sl Proc.\ IEEE\/} ({\sl Lett.\/})}

\def\PRA{{\sl Phys.\ Rev.\ A\/}}

\def\PRL{{\sl Phys.\ Rev.\ Lett.}}
\def\PRSLA{{\sl Proc.\ Roy.\ Soc.\ Lond.\ A\/}}

\def\QIP{{\sl Quantum Inform.\ Processing}}

\def\SIAMJC{{\sl SIAM J. Comput.}}

\def\dajm{\hbox{D. A. Meyer}}

\def\bv{\hbox{E. Bernstein and U. Vazirani}}

\def\grover{\hbox{L. K. Grover}}

\def\shor{\hbox{P. W. Shor}}

\def\hfb{\hfil\break}

\catcode`@=11
\newskip\ttglue

   \font\ninerm=cmr9    \font\eightrm=cmr8   \font\sixrm=cmr6
  \font\ninebf=cmbx9   \font\eightbf=cmbx8  \font\sixbf=cmbx6
  \font\nineit=cmti9   \font\eightit=cmti8  
  \font\ninesl=cmsl9   \font\eightsl=cmsl8  
  \font\ninemi=cmmi9   \font\eightmi=cmmi8  \font\sixmi=cmmi6

\font\bigten=cmr10 scaled\magstep2 

\def\ninepoint{\def\rm{\fam0\ninerm}%
  \textfont0=\ninerm \scriptfont0=\sixrm
  \textfont1=\ninemi \scriptfont1=\sixmi
  \textfont\itfam=\nineit  \def\it{\fam\itfam\nineit}%
  \textfont\slfam=\ninesl  \def\sl{\fam\slfam\ninesl}%
  \textfont\bffam=\ninebf  \scriptfont\bffam=\sixbf
    \def\bf{\fam\bffam\ninebf}%
  \tt \ttglue=.5em plus.25em minus.15em
  \normalbaselineskip=11pt
  \setbox\strutbox=\hbox{\vrule height8pt depth3pt width0pt}%
  \normalbaselines\rm}

\def\eightpoint{\def\rm{\fam0\eightrm}%
  \textfont0=\eightrm \scriptfont0=\sixrm
  \textfont1=\eightmi \scriptfont1=\sixmi
  \textfont\itfam=\eightit  \def\it{\fam\itfam\eightit}%
  \textfont\slfam=\eightsl  \def\sl{\fam\slfam\eightsl}%
  \textfont\bffam=\eightbf  \scriptfont\bffam=\sixbf
    \def\bf{\fam\bffam\eightbf}%
  \tt \ttglue=.5em plus.25em minus.15em
  \normalbaselineskip=9pt
  \setbox\strutbox=\hbox{\vrule height7pt depth2pt width0pt}%
  \normalbaselines\rm}

\def\sfootnote#1{\edef\@sf{\spacefactor\the\spacefactor}#1\@sf
      \insert\footins\bgroup\eightpoint
      \interlinepenalty100 \let\par=\endgraf
        \leftskip=0pt \rightskip=0pt
        \splittopskip=10pt plus 1pt minus 1pt \floatingpenalty=20000
        \parskip=0pt\smallskip\item{#1}\bgroup\strut\aftergroup\@foot\let\next}
\skip\footins=12pt plus 2pt minus 2pt
\dimen\footins=30pc

\def\ie{{\it i.e.}}
\def\eg{{\it e.g.}}

\def\etal{{\it et al.}}

\def\dist{\hbox{\ss dist}}
\def\sdist{\hbox{\tinyss dist}}
\def\wt{\hbox{\ss wt}}
\def\swt{\hbox{\tinyss wt}}
\def\bone{b_1}
\def\mod{{\rm mod}\ }

\def\Lemma{L{\eightpoint EMMA}}

\def\Theorem{T{\eightpoint HEOREM}}

\def\Corollary{C{\eightpoint OROLLARY}}
\def\Definition{D{\eightpoint EFINITION}}

\def\Proposition{P{\eightpoint ROPOSITION}}
\def\endproof{\vrule height6pt width4pt depth2pt}

\def\and{{\eightpoint AND}}

\def\Htwon{(\C^2)^{\otimes n}}

\def\HunzikerMeyer{1}
\def\Angluinsurvey{2}
\def\AngluinSlonim{3}
\def\SloanTuran{4}
\def\Shor{5}
\def\HSPsurvey{6}
\def\Grover{7}
\def\HMPPR{8}
\def\BernsteinVazirani{9}
\def\BargZhou{10}
\def\Helstrom{11}
\def\YKL{12}
\def\Kennedy{13}
\def\Helstrombook{14}
\def\Plotkin{15}
\def\Deutsch{16}
\def\DeutschJozsa{17}
\def\Jozsa{18}
\def\CEMM{19}
\def\Cauchy{20}
\def\SKIH{21}

\magnification=1200

\dimen0=\hsize \divide\dimen0 by 13 \dimendef\chasm=0
%\dimen1=\hsize \advance\dimen1 by -\chasm \dimendef\usewidth=1
%\dimen2=\usewidth \divide\dimen2 by 2 \dimendef\halfwidth=2
%\dimen3=\usewidth \divide\dimen3 by 3 \dimendef\thirdwidth=3
%\dimen4=\hsize \advance\dimen4 by -\halfwidth \dimendef\secondstart=4
%\dimen5=\halfwidth \advance\dimen5 by -10pt \dimendef\indenthalfwidth=5
%\dimen6=\thirdwidth \multiply\dimen6 by 2 \dimendef\twothirdswidth=6
%\dimen7=\twothirdswidth \divide\dimen7 by 4 \dimendef\qttw=7
%\dimen8=\chasm \multiply\dimen8 by 3 \dimendef\onepointfivein=8
%\dimen9=\chasm \multiply\dimen9 by 4.5 \dimendef\twopointtwofivein=9
%\dimen10=\chasm \multiply\dimen10 by 10 \dimendef\fivein=10
\dimen1=\chasm \multiply\dimen1 by  6 \dimendef\halfwidth=1
\dimen2=\chasm \multiply\dimen2 by  7 \dimendef\secondstart=2
\dimen3=\chasm \divide\dimen3 by 2 \dimendef\quarter=3
\dimen4=\quarter \multiply\dimen4 by 9 \dimendef\twopointtwofivein=4
\dimen5=\chasm \multiply\dimen5 by 3 \dimendef\onepointfivein=5
\dimen6=\chasm \multiply\dimen6 by 7 \dimendef\threepointfivein=6
\dimen7=\hsize \advance\dimen7 by -\chasm \dimendef\usewidth=7
\dimen8=\chasm \multiply\dimen8 by 4 \dimendef\thirdwidth=8
\dimen9=\usewidth \divide\dimen9 by 2 \dimendef\halfwidth=9
\dimen10=\usewidth \divide\dimen10 by 3 
                   \multiply\dimen10 by 2 \dimendef\twothirdswidth=10

\parskip=0pt\parindent=0pt

\line{\hfil 2 December 2009}
\vfill
\centerline{\bf\bigten SINGLE-QUERY LEARNING}
\smallskip
\centerline{\bf\bigten FROM ABELIAN AND 
                       NON-ABELIAN$\vphantom{\hbox{\bf\bigten Q}}$}
\smallskip
\centerline{\bf\bigten HAMMING DISTANCE ORACLES}
\bigskip\bigskip
\centerline{\bf David A. Meyer$^*$ and James Pommersheim$^{*,\dagger}$}
\bigskip 
\centerline{\sl $^*$Project in Geometry and Physics,
                Department of Mathematics}
\centerline{\sl University of California/San Diego,
                La Jolla, CA 92093-0112}
\smallskip
\centerline{\sl $^{\dagger}$Department of Mathematics}
\centerline{\sl Reed College, Portland, OR 97202-8199}
\smallskip
\centerline{{\tt dmeyer@math.ucsd.edu},
            {\tt jamie@reed.edu}}
            
\smallskip

\vfill
\centerline{ABSTRACT}
\bigskip
%--------|---------|---------|---------|---------|---------|---------|
\noindent We study the problem of identifying an $n$-bit string using 
a single quantum query to an oracle that computes the Hamming distance 
between the query and hidden strings.  The standard action of the 
oracle on a response register of dimension $r$ is by powers of the
cycle $(1\ldots r)$, all of which, of course, commute.  We introduce a 
new model for the action of an oracle---by general permutations in
$S_r$---and explore how the success probability depends on $r$ and on 
the map from Hamming distances to permutations.  In particular, we 
prove that when $r = 2$, for even $n$ the success probability is $1$ 
with the right choice of the map, while for odd $n$ the success 
probability cannot be $1$ for any choice.  Furthermore, for small odd 
$n$ and $r = 3$, we demonstrate numerically that the image of the 
optimal map generates a non-abelian group of permutations.

\bigskip\bigskip
%--------|---------|---------|---------|---------|---------|---------|
\noindent 2010 Physics and Astronomy Classification Scheme:
                   03.67.Ac. % Quantum algorithms, protocols, and
                             %  simulations

\noindent 2010 American Mathematical Society Subject Classification:
                   68Q12,    % Computer science
                             %  Theory of computing 
                             %   Quantum algorithms and complexity
                   68Q32,    % Computer science
                             %  Theory of computing
                             %   Computational learning theory
	               05B20.    % Combinatorics
                             %  Designs and configurations
                             %   Matrices (incidence, Hadamard, etc.)

\smallskip
\global\setbox1=\hbox{Key Words:\enspace}
\parindent=\wd1
\item{Key Words:}  Quantum algorithms, permutation model.   

\vfill
\eject

\headline{\ninepoint\it Learning from Hamming distance oracles 
                                         \hfill  Meyer \& Pommersheim}

\parskip=10pt
\parindent=20pt

\noindent{\bf 1.  Introduction}

%--------|---------|---------|---------|---------|---------|---------|
Suppose we wish to identify an $n$-bit string $a$ by querying an 
oracle that computes the Hamming distance of any query $x$ from $a$.
Previous work has shown that if the oracle returns the Hamming 
distance modulo 4, there is a quantum algorithm that identifies $a$
with probability 1, using only a single query [\HunzikerMeyer].  On 
the other hand, if the oracle returns the Hamming distance modulo 2, 
there is no algorithm, either classical or quantum mechanical, that 
can identify $a$ with probability greater than $1/2^{n-1}$, using any
number of queries.%
\sfootnote{$^1$}{This follows from the fact that the weight of $a$ 
modulo 2 partitions the set of $n$-bit strings into two subsets of 
size $2^{n-1}$, with each element having even Hamming distance from 
the elements in the same subset and odd Hamming distance from the 
elements in the other subset.}
In the latter case, we can think of the oracle adding the Hamming
distance into a two dimensional response register (so its remainder 
modulo 2 is all that matters), or we can think of the oracle adding
a single bit---the least significant bit of the Hamming 
distance---into a two dimensional response register.  The latter point 
of view might lead us to believe that the difficulty stems from the
oracle returning only a single bit, compared to the two bits that it 
returns when it computes the Hamming distance modulo 4.

%--------|---------|---------|---------|---------|---------|---------|
Our first, possibly surprising, result demonstrates that when $n$ is 
even this belief is wrong---there is a quantum algorithm that takes a 
single bit from the Hamming distance computed by the oracle and 
identifies $a$ with probability 1 using a single query.  Knowing such
an algorithm exists, our second result is perhaps equally surprising:
when $n$ is odd the original belief is at least partially
correct---there is no probability 1 algorithm for finding $a$ using 
any single bit of the Hamming distance.  By ``any single bit of the 
Hamming distance'' we mean any function 
$$
g_a(x) = h\bigl(\dist(a,x)\bigr),                           \eqno(1.1)
$$ 
where $h : \{0,\ldots,n\}\to\{0,1\}$.  Both of these results involve
{\sl learning\/} (or failing to {\sl learn\/}) an element from a set 
of binary functions of $x$, indexed by $a$, so they can be understood 
as solutions to problems in {\sl computational learning theory\/} 
where the set is a {\sl concept class\/} and its elements are 
{\sl concepts\/} [\Angluinsurvey].

%--------|---------|---------|---------|---------|---------|---------|
Combining our new results with the previous ones leads us to make two
observations:  the probability of correctly learning $a$ depends on
(1) the dimension of the response register and (2) how the oracle's 
response acts on this register.  The first observation suggests 
generalizing the notion of concepts, which are binary functions, to 
$Y$-valued functions, for sets $Y$ other than $\{0,1\}$.%
\sfootnote{$^2$}{Limited versions of this generalization have 
been considered previously.  See, for example, 
[\AngluinSlonim,\SloanTuran].}
The second observation motivates the main conceptual contribution of
this paper---a new model for the action of quantum (and reversible
classical) {\sl non-abelian\/} oracles---the permutation model.  In 
this model, we fix a response register $\C^R$, where $R$ is a finite 
set, and assign to each possible reponse $y \in Y$ a permutation 
$\sigma_y\in S_R$ of the set $R$.  More precisely, to implement an 
oracle which computes the (classical) function $f:X\rightarrow Y$, we 
are free to choose any map $\sigma:Y\rightarrow S_R$, and given this 
choice, the oracle acts on $\C^X\otimes \C^R$ by
$$
{\cal O}(f)|x,b\rangle = |x,\sigma_{\!f(x)}(b)\rangle                
$$
When the oracle acts in the standard way, by adding the function 
value it computes into the response register, 
$\sigma(Y)\subseteq C_r \le S_R$, where $C_r$ is the cyclic group with
$r = |R|$ elements, and is thus abelian.  But this need not be the 
case:  $\sigma(Y)$ can generate a non-abelian subgroup of $S_R$ when 
$r > 2$, and for some problems the optimal solution has this property.

%--------|---------|---------|---------|---------|---------|---------|
Our final set of results addresses the problem of {\sl maximizing\/} 
the probability of success for odd $n$ within this permutation model.  
We emphasize that although in this paper we study only Hamming 
distance oracles, {\sl any\/} non-trivial oracle can be set up to have 
a non-abelian action, and this can improve the probability of success 
relative to an abelian action, as it does for the oracles we consider.

\medskip
\noindent{\bf 2.  Background}

\nobreak
%--------|---------|---------|---------|---------|---------|---------|
Most quantum algorithms include one or more calls to a subroutine or
oracle that evaluates some function at the argument passed to it.  In 
some cases, like Shor's algorithm [\Shor] and the various quantum 
algorithms for hidden subgroup problems [\HSPsurvey], the range of 
this function is a large set $Y$ (so that, for example, the function 
can take distinct values on distinct cosets of the hidden subgroup).  
In others, like Grover's algorithm [\Grover], the range of the 
function is only $\{0,1\}$.

%--------|---------|---------|---------|---------|---------|---------|
In the latter cases, the problem of identifying the function can be
recognized as a problem in computational learning theory [\HMPPR]:  
The set of possible functions ${\cal C}\subseteq\{0,1\}^X$, where $X$ 
is the domain of the function, is the {\sl concept class\/}; each 
function $c : X \to \{0,1\}$ is a {\sl concept}; and 
$c^{-1}(1)\subseteq X$ is the {\sl extension\/} of the concept $c$.  
{\sl Concept learning\/} is the process by which a student (the 
learner) identifies (or approximates) a target concept $\bar c$ from a 
concept class ${\cal C}$.  In active learning the student can query a 
teacher for information about the target concept.  Asking a teacher if 
$x\in X$ is in the extension of $\bar c$ is equivalent to passing $x$ 
to a subroutine or oracle that evaluates $\bar c$ at its argument.

%--------|---------|---------|---------|---------|---------|---------|
Many natural concept learning problems---including Grover's [\Grover]
U{\eightpoint NSTRUCTURED} S{\eightpoint EARCH} problem;
Bernstein and Vazirani's [\BernsteinVazirani], and Barg and Zhou's
[\BargZhou], S{\eightpoint IMPLEX} C{\eightpoint ODE} 
D{\eightpoint ECODING} problem; and Hunziker, {\it et al}.'s [\HMPPR] 
B{\eightpoint ATTLESHIP} and M{\eightpoint AJORITY} problems---are 
highly symmetric.  In each of these $|{\cal C}| = |X|$ and there is an 
abelian group $G$ action on ${\cal C}$ and $X$ that is transitive and 
satisfies $(g\cdot c)(g\cdot x) = c(x)$ for all $c\in{\cal C}$, 
$x\in X$, and $g\in G$.

%--------|---------|---------|---------|---------|---------|---------|
In this paper we consider problems which have this symmetry for 
$G = X = \Z_2^n$.  Each involves a specific function of the Hamming 
distance between some unknown $n$-bit string $a\in\Z_2^n$ and 
$x\in\Z_2^n$, $\dist(a,x) = |\{i\mid a_i \not= x_i\}|$; this is 
invariant under the action of $G$ since $\dist(g+a,g+x) = \dist(a,x)$.  
Now, until it is composed with a binary function as in (1.1), 
$\dist(a,\cdot) : X \to \{0,\ldots,n\} = Y$ does not define a 
traditional concept (except in the trivial case $n = 1$), so it is
useful to define a $Y$-{\sl valued concept class\/} to be a set of 
functions ${\cal C}\subseteq Y^X$.  We extend our use of ``learning 
problems'' to include these cases.

%--------|---------|---------|---------|---------|---------|---------|
\noindent\Definition.  An $(n,r)$-{\sl Hamming distance oracle\/} 
accepts queries $x\in\Z_2^n$ and then acts on an $r$-dimensional 
response register according to some function of the Hamming distance 
$\dist(a,x)$, for some fixed $a\in\Z_2^n$.%
\sfootnote{$^3$}{It is also natural to consider problems with 
$X = \sZ_k^n$ [\HunzikerMeyer], in which the Hamming distance is 
defined by the same formula, but in this paper we restrict our 
attention to $k = 2$.}

%--------|---------|---------|---------|---------|---------|---------|
Our goal is to optimize single-query learning from such Hamming 
distance oracles, \ie, to maximize the probability of correctly 
identifying $a$ after a single call to the subroutine that computes 
the function.  Since we assume a uniform distribution on $a$, we 
consider only quantum algorithms that begin with an {\sl equal 
superposition query},%
\sfootnote{$^4$}{In fact, we conjecture that for problems with
transitive group actions and uniform priors, the optimal solutions 
always include one that begins with an equal superposition query.}
\ie, that pass to the oracle a state of the form 
$|\eta^0\rangle\otimes\psi = 
 H^{\otimes n}|0\ldots0\rangle\otimes\psi$, where $H$ is the Hadamard 
transformation $\bigl({1\atop 1}{1\atop -1}\bigr)/\sqrt{2}$ and 
$\psi\in\C^r$.  If 
${\cal O}(a) : (\C^2)^{\otimes n}\otimes\C^r 
               \to 
               (\C^2)^{\otimes n}\otimes\C^r$ denotes the action of 
the oracle with parameter $a$, the problem reduces to identifying 
which of the $2^n$ states ${\cal O}(a)|\eta^0\rangle\otimes\psi$ is 
returned by the oracle.  An optimal solution to this problem can be 
obtained by a complete von Neumann measurement
[\Helstrom,\YKL,\Kennedy,\Helstrombook]; equivalently, we want to 
maximize
$$
\sum_{a=0}^{2^n-1}\sum_{b=0}^{r-1} 
 \Bigl|\langle a,b|U{\cal O}(a)|\eta^0\rangle\otimes\psi\Bigr|^2,
                                                            \eqno(2.1)
$$ 
over all unitary maps $U\in U(2^nr)$ and states $\psi\in\C^r$.

\medskip
\noindent{\bf 3.  Using a different bit of the Hamming distance}

\nobreak
%--------|---------|---------|---------|---------|---------|---------|
We begin by considering the problem of learning an $n$-bit string from
an oracle that returns the {\sl second\/} least significant bit of the
Hamming distance of a query, rather than the least significant bit as
in [\HunzikerMeyer].  To be precise, let $n$ be a natural number, and 
for any $a\in\Z_2^n$, define a function 
$f_a : \Z_2^n \to \{0,1\}$ by
$$
f_a(x) = 
 \cases{0 & if $\dist(a,x)\equiv 0,1$ (mod $4$);           \cr
        1 & if $\dist(a,x)\equiv 2,3$ (mod $4$).           \cr
       }
$$
Thus $f_a(x)$ is the second least significant bit of the Hamming 
distance between $a$ and $x$.  Set $\bone(d)$ to be the second least 
significant bit of a nonnegative integer $d$, so 
$f_a(x)=\bone\bigl(\dist(a,x)\bigr)$.  Define ${\cal C}_n$ to be the 
concept class $\{f_a \mid a\in\Z_2^n\}$.

%--------|---------|---------|---------|---------|---------|---------|
\noindent\Lemma\ 3.1.  {\sl If $n \not\equiv 1$ {\rm (mod $4$)} then 
$|{\cal C}_n| = 2^n$.  If $n \equiv 1$ {\rm (mod $4$)} then
$f_a = f_{\bar a}$, where $\bar a$ is the bitwise complement of 
$a\in\Z_2^n$, so there are only $2^{n-1}$ concepts in the class.}

%--------|---------|---------|---------|---------|---------|---------|
\noindent{\sl Proof}.  Suppose $f_{a'} = f_a$ and $\dist(a,a') = d$.
Since 
$\bone(d) = \bone\bigl(\dist(a',a)\bigr) = f_{a'}(a) = f_a(a) 
 = \bone\bigl(\dist(a,a)\bigr) = \bone(0) = 0$, 
we must have $d \equiv 0$ or $1$ (mod $4$).  If $a'\not=a$ there is a 
bit at which $a'$ differs from $a$.  Let $x$ be the bit string 
obtained from $a$ by complementing this bit.  Then 
$\bone\bigl(\dist(a,x)\bigr) = \bone(1) = 0$ so 
$\bone\bigl(\dist(a',x)\bigr) = \bone(d-1) = 0$, so we can conclude 
that $d \equiv 1$ (mod $4$).  Now suppose there were a bit at which 
$a'$ agreed with $a$.  Let $y$ be the bit string obtained from $a$ by
complementing this bit.  Then 
$\bone\bigl(\dist(a,y)\bigr) = \bone(1) = 1$ and
$\bone\bigl(\dist(a',y)\bigr) = \bone(d+1)$, which would imply that
$d \equiv 0$ (mod $4$), a contradiction.  So if $a' \not= a$ but
$f_{a'} = f_a$, there can be no bit at which $a'$ agrees with $a$, 
which means $a' = \bar a$ and $n\equiv 1$ (mod $4$).\hfill\endproof

%--------|---------|---------|---------|---------|---------|---------|
As we explained in the previous sections, we are interested in 
analyzing the probability of correctly identifying the hidden bit 
string $a$ using only a single query to the oracle.  Classically, it 
is not hard to see that when the $f_a$ are distinct, any strategy 
yields a worst-case success probability of at most 
$2/2^n = 1/2^{n-1}$, the number of possible oracle responses divided 
by the number of concepts.  In contrast, we next show that for even 
$n$, this learning problem can be solved quantum mechanically with 
probability 1 using a single query.

%--------|---------|---------|---------|---------|---------|---------|
\noindent\Theorem\ 3.2.  {\sl Let $n$ be even.  Then the 
learning problem defined by ${\cal C}_n$ can be solved with 
probability 1 using a single quantum query.}

%--------|---------|---------|---------|---------|---------|---------|
We will prove Theorem 3.2 by giving an explicit algorithm below.  To 
show that the algorithm is correct we will need two lemmas.  For 
$x \in \Z_2^n$, define $\hat{x} \in \Z_2^n$ by:
$$
\hat{x} = 
  \cases{      x & if $\wt(x)$ is even;   \cr
         \bar{x} & if $\wt(x)$ is odd.    \cr
        }
$$
Here the {\sl weight\/} of $x$, $\wt(x) = \dist(0,x)$.  Note that if 
$n$ is even, then the function $x \mapsto \hat{x}$ is a permutation of 
$\Z_2^n$.

%--------|---------|---------|---------|---------|---------|---------|
\noindent\Lemma\ 3.3. {\sl Let $n$ be a natural number and let 
$a,x \in \Z_2^n$.  Then} 
$$
a \cdot x +\wt(a)\wt(x) \equiv  a \cdot \hat{x}\ (\mod 2).
$$

%--------|---------|---------|---------|---------|---------|---------|
\noindent{\sl Proof}.  If $\wt(x)$ is even, then $\hat{x}=x$, and the 
congruence is easily seen to hold.  If $\wt(x)$ is odd, then 
$\hat{x}=\bar{x}$, and the congruence follows from the identity
$a \cdot x + a \cdot \bar{x} = \wt(a)$.                \hfill\endproof

%--------|---------|---------|---------|---------|---------|---------|
\noindent\Lemma\ 3.4. {\sl Let $n$ be a natural number and let 
$a,x \in \Z_2^n$. Then} 
$$ 
(-1)^{\bone(\sdist(a,x))} 
 = 
(-1)^{\bone(\swt(a))}(-1)^{\bone(\swt(x))}(-1)^{a \cdot\hat{x}}.
$$

%--------|---------|---------|---------|---------|---------|---------|
\noindent{\sl Proof}.  First note that for any integer $d$,
$$
\bone(d) \equiv {{d(d-1)} \over 2}\quad (\mod 2).
$$
Since $\dist(a,x) = \wt(a) + \wt(x) - 2(a\cdot x)$, this implies
$$
\bone\bigl(\dist(a,x)\bigr) 
 \equiv 
{\bigl(\wt(a) + \wt(x) - 2(a\cdot x)\bigr)
 \bigl(\wt(a) + \wt(x) - 2(a\cdot x) - 1\bigr)
 \over 2
}\quad (\mod 2).
$$
Expanding the numerator on the right hand side of this congruence, and 
dropping multiples of $4$, gives
$$
\bone(\dist(a,x)) 
 \equiv 
{\wt(a)^2 - \wt(a) + \wt(x)^2 - \wt(x) 
 \over 
 2
} + \wt(a)\wt(x) + a \cdot x\quad (\mod 2).
$$
Using Lemma 3.3, we can replace $\wt(a)\wt(x) + a \cdot x$ with 
$a \cdot \hat{x}$ and the result follows.              \hfill\endproof

%--------|---------|---------|---------|---------|---------|---------|
\noindent{\sl Proof of Theorem 3.2}.  We take the oracle to act on 
$\Htwon\otimes\C^2$ in the standard way, 
$$
{\cal O}(a) : |x\rangle|b\rangle \mapsto |x\rangle|b+f_a(x)\rangle,
$$
although it is $\bone\bigl(\dist(a,x)\bigr)$ that is being added into
the response register, not $\dist(a,x)$.  The following quantum 
algorithm identifies $a$ with probability $1$, applying ${\cal O}(a)$ 
only once.
\vskip6pt

\parskip=3pt
\noindent{\ssbf Algorithm A}.
\smallskip
\item{{\ssbf 1}.} 
%--------|---------|---------|---------|---------|---------|---------|
Initialize the state to 
$|0\ldots0\rangle|0\rangle \in \Htwon\otimes\C^2$.

\item{{\ssbf 2}.}
%--------|---------|---------|---------|---------|---------|---------|
Apply the unitary transformation $H^{\otimes n} \otimes HX$, where 
$X = \bigl({0\atop 1}{1\atop 0}\bigr)$.  This produces the state
$$
|\eta^0\rangle|-\rangle
 = 
{1\over2^{n/2}} \sum_{x\in \sZ_2^n} |x\rangle|-\rangle,
$$
where $|-\rangle = (|0\rangle - |1\rangle)/\sqrt{2}$.

\item{{\ssbf 3}.}  
%--------|---------|---------|---------|---------|---------|---------|
Let $D$ be the diagonal matrix acting on $\Htwon$ by 
$D|x\rangle = (-1)^{\bone(\swt(x))}|x\rangle$.   Apply $D\otimes I$,
producing the state
$$
{1\over2^{n/2}} \sum_{x\in \sZ_2^n} 
 (-1)^{\bone(\swt(x))}|x\rangle |-\rangle.
$$

\item{{\ssbf 4}.}  
%--------|---------|---------|---------|---------|---------|---------|
Apply the oracle ${\cal O}(a)$.  This produces the state
$$
{1\over2^{n/2}} \sum_{x\in\sZ_2^n}(-1)^{\bone(\swt(x))} 
(-1)^{\bone(\sdist(a,x))} |x\rangle|-\rangle.
$$
By Lemma 3.4, this equals
$$
{1\over2^{n/2}} \sum_{x\in\sZ_2^n} 
(-1)^{\bone(\swt(a))}(-1)^{a \cdot \hat{x}}|x\rangle|-\rangle.
$$

\item{{\ssbf 5}.}
%--------|---------|---------|---------|---------|---------|---------|
Let $P$ be the permutation matrix acting on $\Htwon$ by 
$P|x\rangle = |\hat{x}\rangle$.   Applying $P\otimes I$ yields
$$
{1\over2^{n/2}} \sum_{x\in \sZ_2^n} 
(-1)^{\bone(\swt(a))}(-1)^{a \cdot \hat{x}}|\hat{x}\rangle|-\rangle,
$$
which is equal to
$$
{(-1)^{\bone(\swt(a))}\over2^{n/2}} 
\sum_{x\in\sZ_2^n} (-1)^{a \cdot x}|x\rangle|-\rangle,
$$
since $x\mapsto\hat{x}$ is a bijection.

\item{{\ssbf 6}.}
%--------|---------|---------|---------|---------|---------|---------|
Apply $H^{\otimes n} \otimes I$.  This produces the state
$(-1)^{\bone(\swt(a))} |a\rangle|-\rangle$.

\item{{\ssbf 7}.}
%--------|---------|---------|---------|---------|---------|---------|
Now measure the query register (the $\Htwon$ tensor factor) and 
observe $a$ with probability 1.                        \hfill\endproof

\parskip=10pt
\medskip
\noindent{\bf 4.  Concept classes that cannot be learned with a single
                  query}

\nobreak
%--------|---------|---------|---------|---------|---------|---------|
Theorem 3.2 cannot be extended to odd $n > 1$; there is no single 
equal superposition query probability 1 quantum learning algorithm for 
the concept class ${\cal C}_n$ in this case.  In fact, when $n > 1$ is 
odd, there is no concept class defined by any function of the Hamming 
distance that is perfectly learnable with a single equal superposition 
quantum query.  To see this, we begin with the following lemma:  

%--------|---------|---------|---------|---------|---------|---------|
\noindent\Lemma\ 4.1.  {\sl Let ${\cal C}$ be a concept class of size 
$M$ over a set $X$ of size $N$.  Suppose that there is a probability 
$1$ learning algorithm using a single equal superposition quantum 
query.  Identifying concepts with bitstrings indexed by $X$, there 
exists an integer $d\geq N/2$ such that any two distinct concepts of 
${\cal C}$ have Hamming distance $d$.  If $M = N > 2$ is even, then 
the quantum learning matrix, which has entries $L_{xc} = (-1)^{c(x)}$
for $x\in X$, $c\in{\cal C}$, is a Hadamard matrix.}

%--------|---------|---------|---------|---------|---------|---------|
\noindent{\sl Proof}.  Suppose that there is a single query learning 
algorithm with equal superposition query
$$
{1\over \sqrt{N}} \sum_{x\in X} |x\rangle \otimes \psi,
$$
for some unit vector $\psi\in\C^2$.  If 
$\lambda = \psi^{\dagger}X\psi$ then $-1\le\lambda\le1$.%
\sfootnote{$^5$}{This $X$ is the bit-flip matrix defined in step 2 of 
Algorithm A, not the set over which the concept class is defined.}
Let $A$ be the matrix whose columns, indexed by concepts, contain the 
state of the system after querying the oracle.  Then $B=A^{\dagger}A$ 
is a matrix whose rows and columns are both indexed by concepts, with 
elements
$$
B_{cc'} = {1\over N}
\sum_{x\in X}\cases{      1 & if $c(x) = c'(x)$;           \cr
                    \lambda & if $c(x) \neq c'(x)$.        \cr
                   }
$$
Thus $NB_{cc'} = \bigl(N-\dist(c,c')\bigr)+\lambda\,\dist(c,c')$.  
Since the algorithm succeeds with probability 1, we must have 
$B_{cc'} = 0$ for distinct concepts $c\neq c'$.  In this case
$$
d = \dist(c,c') = {N \over 1-\lambda} \ge {N\over 2},
$$
where the inequality follows from $\lambda \geq -1$.  

%--------|---------|---------|---------|---------|---------|---------|
Now suppose that $M = N > 2$ is even.  Note that the the concepts of 
${\cal C}$ form a code of distance $d$.  Hence if $d > N/2$, then 
the Plotkin bound [\Plotkin] implies that 
$$
M\leq 2 \biggl\lfloor  {d\over 2d - N} \biggr\rfloor. 
$$  
Since $N$ is even, $2d - N \ge 2$, and it follows that $M < N$ unless 
$d = N$, in which case $M \le 2$.  Thus we must have $d = N/2$ so the 
columns of $L$ are orthogonal.  That is, if $M = N > 2$ is even, the 
quantum learning matrix is a Hadamard matrix.          \hfill\endproof

%--------|---------|---------|---------|---------|---------|---------|
We now use Lemma~4.1 to prove the general result:

%--------|---------|---------|---------|---------|---------|---------|
\noindent\Theorem\ 4.2.  {\sl Let $n > 1$ be odd.  Suppose that  
${\cal E}_n = \{g_a \mid a\in\Z_2^n\}$,  where the functions 
$g_a : \Z_2^n \to \Z_2^{\vphantom n}$ have the property that $g_a(x)$ 
depends only on the Hamming distance $\dist(a,x)$.  If 
$|{\cal E}_n| = 2^n$, then the learning problem defined by 
${\cal E}_n$ cannot be solved with probability 1 using a single 
quantum query.} 
 
%--------|---------|---------|---------|---------|---------|---------|
Note that if $|{\cal E}_n| \neq 2^n$, then $a$ is not determined by 
$g_a$.  Thus, in general, when $n$ is odd, the bitstring $a$ cannot be 
learned with probability 1 in a single quantum query from any 
binary-valued function of the Hamming distance.

%--------|---------|---------|---------|---------|---------|---------|
\noindent{\sl Proof}.  Since $g_a(x)$ depends only on the Hamming 
distance $\dist(a,x)$, there exists a function 
$h:\{0,\dots,n\}\to \{0,1\}$ such that 
$g_a(x) = h\bigl(\dist(a,x)\bigr)$.

%--------|---------|---------|---------|---------|---------|---------|
Suppose that the learning problem defined by ${\cal E}_n$ can be 
solved with probability $1$ using a single quantum query.  Then by 
Lemma~4.1, the quantum learning matrix $L$, with elements 
$L_{xa} = (-1)^{g_a(x)}$, is a Hadamard matrix.  Consider the inner 
product of the two rows of $L$ corresponding to the queries $y = 0^n$ 
and $z = 1^20^{n-2}$.  Since $L$ is a Hadamard matrix, 
$$
\sum_{a\in\sZ_2^n} (-1)^{g_a(y)} (-1)^{g_a(z)} = 0.
$$
In half of the terms of this sum, those for which the bits $a_0$ and 
$a_1$ differ, $\dist(a,y) = \dist(a,z)$.  Then $g_a(y) = g_a(z)$, and 
hence each of these terms contributes $+1$ to the sum.  In the other 
half of the terms, those for which $a_0 = a_1$, each term must 
contribute $-1$ to the sum, so 
$g_a(y) \equiv g_a(z) + 1\,(\mod 2)$.  But $a_0 = a_1$ implies 
$\dist(a,y) = \dist(a,z) \pm 2$.  It follows that for any 
$d\in\{0,\dots,n-1\}$, $h(d)\neq h(d+2)$.  Hence for some 
$s\in\{0,1,2,3\}$, 
$
h(d) = \bone(d+s)
$
for all $d\in\{0,\dots,n\}$.  Thus under the assumption that the 
concept class can be learned with probability 1 from a single quantum 
query, we have shown that $h$ is a translate of $b_1$. 

%--------|---------|---------|---------|---------|---------|---------|
It remains to show that if $n$ is odd, taking $h$ to be a translate of 
$b_1$ leads to a matrix $L$ that is not a Hadamard matrix.  One easily 
sees that for such a function $h$, there is a sign $\epsilon = \pm1$ 
such that 
$$
(-1)^{h(n-d)} =\epsilon (-1)^{h(d)}
$$
for all $d$.   It follows that any two rows of $L$ corresponding to 
complementary values of $x$ are equal up to sign.  Hence $L$ is not a 
Hadamard matrix.                                       \hfill\endproof

%--------|---------|---------|---------|---------|---------|---------|
When $n\equiv 3$ (mod $4$), the concept class ${\cal C}_n$ we 
introduced in the previous section satisfies the hypotheses of 
Theorem~4.2, so it cannot be learned with probability $1$ from a 
single quantum query.  When $n\equiv 1$ (mod $4$), Lemma~3.1 tells us 
that the concept class has only $2^{n-1}$ concepts so Theorem~4.2 does 
not apply to learning the concept classe ${\cal C}_n$ in this case.  We
already know in this case that $a$ cannot be identified with 
probability greater than $1/2$ with any number of queries, since 
$f_a = f_{\bar a}$.  Using Algorithm A (with appropriate minor 
modifications), however, a single query determines $a$ up to 
complementation, so the {\sl concept class\/} ${\cal C}_n$  can be 
learned with a single quantum query.

%--------|---------|---------|---------|---------|---------|---------|
Notice that we did not use the fact that $n$ is odd to reach the 
conclusion that $h$ is a translate of $b_1$.  This means that for 
even $n$, the Hamming distance concept class ${\cal C}_n$ is 
essentially the only one that can be learned with probability $1$ 
using a single query.  More precisely, we have:

%--------|---------|---------|---------|---------|---------|---------|
\noindent\Corollary\ 4.3.  {\sl When $n$ is even, $b_1$} ({\sl and 
translates\/}) {\sl are the only functions of Hamming distance that 
yield a concept class learnable with probability $1$ using a single 
quantum query.}

\medskip
\noindent{\bf 5.  The permutation model}

\nobreak
%--------|---------|---------|---------|---------|---------|---------|
The results of the previous section demonstrate that an $n$-bit string
$a$ cannot be learned with probability $1$ using a single quantum 
query to an $(n,2)$-Hamming distance oracle, when $n$ is odd.  A 
natural question, then, is:

\itemitem{}
%--------|---------|---------|---------|---------|---------|---------|
What is the largest probability with which $a$ can be learned using a 
single quantum query to an $(n,2)$-Hamming distance oracle?

%--------|---------|---------|---------|---------|---------|---------|
\noindent Furthermore, although previous work has shown that $a$ can 
be learned with probability $1$ from an $(n,4)$-Hamming distance 
oracle [\HunzikerMeyer], neither that work nor our results to this 
point address the potential for learning with a $3$-dimensional 
response register.  So there is a second natural question:

\itemitem{}
%--------|---------|---------|---------|---------|---------|---------|
What is the largest probability with which $a$ can be learned using a 
single quantum query to an $(n,3)$-Hamming distance oracle?

%--------|---------|---------|---------|---------|---------|---------|
\noindent Before answering these questions, we reconsider the 
formulation of oracle algorithms.

%--------|---------|---------|---------|---------|---------|---------|
To allow comparison with the classical query complexity of oracle
(learning) problems, the action of the oracle in a quantum algorithm
must be the linear extension of a classical reversible operation.  In
Deutsch's [\Deutsch] and Deutsch and Jozsa's [\DeutschJozsa] original 
quantum algorithms for oracle problems, the oracle acts on 
$\Htwon\otimes\C^2$ by 
$$
{\cal O}(c)|x,b\rangle = |x,b+c(x)\rangle,                  \eqno(5.1)
$$
where the sum is computed modulo $2$, but the second register is 
initialized to $|0\rangle$, so the action has the effect of simply
writing the function value computed by the oracle into that register.
Similarly, in quantum algorithms for hidden subgroup problems [\Jozsa]
the oracle computes a function that is constant on cosets of the 
hidden subgroup, and takes distinct values on distinct cosets, so it 
acts on $\C^N\otimes\C^r$, where $N$ is the size of the group and $r$ 
is the number of distinct cosets, by (5.1), where the sum is computed 
modulo $r$.  Again the second register is initialized to $|0\rangle$ 
so this also has the effect of merely writing the function value into 
that register.

%--------|---------|---------|---------|---------|---------|---------|
Cleve, \etal,%
\sfootnote{$^6$}{And Tapp, according to a note in [\CEMM], and most 
likely others as well}
noticed that the success probability of Deutsch's original algorithm 
could be improved to $1$ by initializing the response register in the 
state $|-\rangle$, thereby taking advantage of the action (5.1) when 
$b = 1$ as well as when $b = 0$, to ``kick back'' a phase of 
$(-1)^{c(x)}$ [\CEMM].  Algorithm A does the same thing.  This 
application, as opposed to the application on the response register 
initialized to $|0\rangle$, emphasizes that ${\cal O}(c)$ acts as a 
map on $\{0,1\}$---the (labels of the) computational basis vectors of 
the $\C^2$ response register---and is a classical reversible operation 
for each of the possible values of $c(x)$:  $0$ acts as the identity 
and $1$ acts to exchange $0$ and $1$.  That is, the oracle response, 
both classically and quantum mechanically, can be thought of as an 
element of $S_2$, the permutations of a two element set---it is either 
the identity, $(1)$, or the other element of $S_2$, the permutation 
$(12)$.  From this point of view, the action of an $(n,2)$-Hamming 
distance oracle depends on a map $\{0,\ldots,n\}\to S_2$:  Simply 
adding the Hamming distance into the response register would be the 
map $d \mapsto (12)^d$, while the Algorithm A oracle action comes from 
the map $d \mapsto (12)^{b_1(d)}$ (using {\sl cycle notation\/} 
[\Cauchy] for permutations of the elements of $R$, which we 
label $\{1,\ldots,r\}$). 

%--------|---------|---------|---------|---------|---------|---------|
But this implies a novel conceptualization of the action of an oracle
when $r > 2$, as it can be for $(n,r)$-Hamming distance oracles, 
namely that the action should depend on a map 
$\sigma : \{0,\ldots,n\}\to S_r$ which takes each function value 
computed by the oracle and associates to it a permutation of a
{\sl response set\/} $R$ with $|R| = r$.  In a quantum algorithm, $R$
is identified with the computational basis of the tensor factor used
as the response register.  The map $\sigma$ can be more complicated 
than $d\mapsto (12\ldots r)^d$, \ie, addition of the Hamming distance 
modulo $r$.  This simple action can be characterized an {\sl abelian
oracle\/} since the range of $\sigma$ is contained in a cyclic 
subgroup of $S_r$.  It allows $a$ to be identified with probability 
$1$ when $r = 4$ [\HunzikerMeyer], but in other cases there is no 
reason to think that it is the optimal action.  In general we should 
consider {\sl non-abelian oracles}, ones for which the range of 
$\sigma$ contains noncommuting permutations of $R$.  More precisely, 
we define the action of an oracle on $\Htwon\otimes\C^r$ by
$$
{\cal O}_{\sigma}(a)
|x,b\rangle = |x,\sigma_{\sdist(a,x)}(b)\rangle,            \eqno(5.2)
$$
and let
$$
p_n(r) 
 = 
\max_{\scriptstyle\sigma : \{0,\ldots,n\}\to S_r
      \atop 
      \scriptstyle  \psi\in\sC^r, U\in U(2^nr)}
\sum_{a=0}^{2^n-1}\sum_{b=0}^{r-1}
 \Bigl|\langle a,b|U{\cal O}_{\sigma}(a)|\eta^0\rangle\otimes\psi
 \Bigr|^2.                                                  \eqno(5.3)
$$
Using this notation, Hunziker and Meyer's result [\HunzikerMeyer] 
shows that $p_n(r) = 1$ for $r\ge4$, Theorem~3.2 shows that 
$p_{2j}(r) = 1$ for $r\ge2$, and Lemma~3.1 and Theorem~4.2 show that 
$p_{2j-1}(2) < 1$, for $j$ any natural number.  Furthermore, the two 
questions above can be phrased as:  What are $p_{2j-1}(2)$ and 
$p_{2j-1}(3)$, respectively?

\medskip
\noindent{\bf 6.  Numerical optimization results}

%--------|---------|---------|---------|---------|---------|---------|
We are considering learning algorithms that send a single equal
superposition query $|\eta^0\rangle\otimes\psi$ to the oracle.  If the 
states $\{{\cal O}(c)|\eta^0\rangle\otimes\psi\mid c\in{\cal C}\}$ are 
linearly independent, then the optimal measurement to distinguish 
them, \ie, to identify $c$, is the {\sl square root measurement}, as 
Sasaki, \etal\ noted [\SKIH] using early results in quantum state
discrimination [\Helstrombook, Appendix A].%
\sfootnote{$^7$}{The introduction of this approach into the context of 
concept learning may be found in [\HMPPR].}
Thus we have the following:

%--------|---------|---------|---------|---------|---------|---------|
\noindent\Proposition\ 6.1.  {\sl Let ${\cal C}$ be a $Y$-valued 
concept class of size $M$ over a set $X$ of size $N$.  Fix a response 
set $R$ and an assignment $\sigma$ of a permutation of $R$ to each 
$y\in Y$.  Also fix the initial state $\psi$ of the response register 
$\C^R$.  Let $B$ be the $Nr$ by $M$ matrix with columns indexed by the 
concepts $c\in{\cal C}$, and with column $c$ the state 
${\cal O}(c)|\eta^0\rangle\otimes\psi$.  Suppose that the columns of 
$B$ are linearly independent, and that the diagonal elements of 
$G = B^\dagger B$ are equal.  Let $\sqrt{G}$ denote the positive 
semi-definite square root of $G$.  Then the optimal single-query 
quantum algorithm using the equal superposition query 
$|\eta^0\rangle\otimes\psi$ succeeds with probability the diagonal 
value in $\sqrt{G}$.}  

%--------|---------|---------|---------|---------|---------|---------|
This proposition justifies the main step in the following numerical
method.
\vskip6pt

\parskip=0pt
\noindent{\ssbf Method B}.
\smallskip
\item{{\ssbf 1}.} 
%--------|---------|---------|---------|---------|---------|---------|
Input $n$ and $r$.

\item{{\ssbf 2}.}
%--------|---------|---------|---------|---------|---------|---------|
Repeat Steps {\ssbf 3} and {\ssbf 4} below for all possible 
assignments $\sigma : \{0,\ldots,n\} \to S_r$.

\itemitem{{\ssbf 3}.}  
%--------|---------|---------|---------|---------|---------|---------|
For $\psi \in \C^r$ a unit vector, Proposition~6.1 allows us to 
calculate the maximal success probability $M(\psi)$ of a single-query 
quantum algorithm using the query $|\eta^0\rangle\otimes\psi$. 

\itemitem{{\ssbf 4}.} 
%--------|---------|---------|---------|---------|---------|---------|
Numerically maximize $M(\psi)$ over all unit vectors $\psi\in\C^r$.

\parskip=10pt
%--------|---------|---------|---------|---------|---------|---------|
\noindent Using this method we obtain the following numerical results:

\item{}
%--------|---------|---------|---------|---------|---------|---------|
First, let $n=3$.

\itemitem{}
%--------|---------|---------|---------|---------|---------|---------|
For $r=2$, we find $p_3(2) \approx 0.800$.  This is achieved using the 
permutations $\sigma_0=\sigma_2=\sigma_3=(1)$ and $\sigma_1=(12)$.  It 
can also be achieved using the permutations 
$\sigma_0=\sigma_1=\sigma_2=(1)$ and $\sigma_3=(12)$.

\itemitem{}
%--------|---------|---------|---------|---------|---------|---------|
When $r=3$, this improves to $p_3(3)\approx 0.974$.  Here a best 
permutation assignment is $\sigma_0=(1)$, $\sigma_1=(12)$, 
$\sigma_2=(132)$, and $\sigma_3=(123)$.  (There are several other 
assignments of permutations that yield the same success probability.)

\item{}
%--------|---------|---------|---------|---------|---------|---------|
Second, let $n=5$.

\itemitem{}
%--------|---------|---------|---------|---------|---------|---------|
When $r=2$, we find $p_5(2)\approx 0.721$.  This is achieved using the 
permutations $\sigma_0=\sigma_3=\sigma_4=\sigma_5=(1)$ and 
$\sigma_1=\sigma_2=(12)$.  

\itemitem{}
%--------|---------|---------|---------|---------|---------|---------|
When $r=3$, this improves to $p_5(3)\approx 0.955$.  Here the best 
permutation assignment is $\sigma_0=(1)$, $\sigma_1=(123)$, 
$\sigma_2=(132)$, $\sigma_3=(12)$, $\sigma_4=(1)$, and 
$\sigma_5=(123)$.  The optimum initialization for the response 
register is approximately
$$
|1\rangle - 0.1065i|2\rangle +1.1064i|3\rangle,
$$
normalized to have unit length. 

%--------|---------|---------|---------|---------|---------|---------|
Note that Proposition 6.1 requires the columns of the matrix $B$ to be 
linearly independent.  In cases that the the columns of $B$ are are 
linearly dependent, Proposition~6.1 does not tell us what to do, and 
Method B may not succeed in finding the optimal solution.  In our 
problem it turns out that certain assignments of permutations lead to 
matrices $B$ with linearly dependent columns.  One suspects that these 
assignments are not as good as the assignments for which $B$ has full 
rank, but this is not guaranteed by Proposition~6.1.  In particular, 
when the rank of $B$ is low, it is generally true that it is 
impossible to distinguish these states with high probability:%
\sfootnote{$^8$}{This is a broadly applicable result that may well 
exist in the literature, but we have been unable to find it.}
%

%--------|---------|---------|---------|---------|---------|---------|
\noindent\Lemma\ 6.2.  {\sl Suppose $\psi_i$, $i\in\{1,\ldots,n\}$ are 
pure states contained in a $k$-dimensional subspace $W$.  Then any 
$n$-valued measurement for identifying $i$ succeeds with probability 
at most $k/n$.}

%--------|---------|---------|---------|---------|---------|---------|
\noindent{\sl Proof}.  Let $\rho_i$ be the density matrix 
corresponding to $\psi_i$.  Let $\Pi_W$ denote projection onto $W$. 
Then $\rho_i \le \Pi_W$ for all $i$.  Hence, if $\{X_i\}$ is any 
measurement, we can bound the success probability of this measurement 
as follows:
$$
{1\over n}\sum_{i=1}^n \Tr(X_i\rho_i) 
 \le 
{1\over n}\sum_{i=1}^n \Tr(X_i\Pi_w) 
 =
{1\over n}\Tr(\Pi_W)
 =
{k \over n}.                                            \eqno\endproof
$$

%--------|---------|---------|---------|---------|---------|---------|
Lemma~6.2 suffices to guarantee that cases in which $B$ has linearly 
dependent columns yield success probabilities that are smaller than 
the ones presented in the list above.  When $n=3$ (for both $r=2$ and 
$r=3$), we find that a given assignment of permutations either leads 
to a matrix $B$ that is full rank (rank $8$) for a generic choice of 
$\psi$, or has rank at most $5$.  In this latter case, Lemma~6.2 
implies that the success probability is at most $5/8$, which is 
smaller that the probabilities shown above in the full rank case.  
When $n=5$, $B$ has either full rank (rank $32$), or rank at most 
$22$, which implies a success probability of at most $22/32$ in the 
linearly dependent case.  Again, this is smaller than the numbers 
reported above for the linearly independent case.

\medskip
\noindent{\bf 7.  Conclusions}

%--------|---------|---------|---------|---------|---------|---------|
We have introduced a novel generalization for the action of oracles in
quantum (and reversible classical) algorithms:  the permutation model.
For $n$-bit Hamming distance oracles the action is specified by a 
choice of map $\sigma:\{0,\ldots,n\}\to S_r$ when the response 
register has dimension $r$.  The standard additive action of the 
oracle is described by the map $\sigma(d) = (1\ldots r)^d$.  Algorithm 
A in Theorem~3.2 demonstrates the striking improvement possible by an 
oracle that acts by some other map of Hamming distances to 
permutations:  for $r = 2$ the success probability of learning from a 
single query to an oracle that acts by the additive action is 
$1/2^{n-1}$, while for any even $n$ it is $1$ for an oracle that acts 
by $\sigma(d) = (12)^{b_1(d)}$, and for $n = 3$ and $n = 5$ it is 
approximately $0.800$ and $0.721$, respectively, using the actions 
listed in \S6.

%--------|---------|---------|---------|---------|---------|---------|
Allowing a larger response register, namely $r =3$, improves the 
latter two probabilities to approximately $0.974$ and $0.955$, 
respectively.  In general, $p_n(r)$ is a nondecreasing function of 
$r$.  One might guess that if there is enough room in the response 
register to encode each possible function value $y\in Y$ as a distinct 
permutation of $\{1,\ldots,r\}$, then adding additional dimensions to 
the response register would not improve the success probability.  This 
guess would mean that $p_n(r)$ would be constant for $r! \ge n+1$.  
This is not the case, however, as the $n = 3$ results show:  
$p_3(3) < 1$ while $p_3(4) = 1$.  

%--------|---------|---------|---------|---------|---------|---------|
As this counterexample indicates, the permutation model raises a host
of new questions.  We close by listing a few more:  Is there some
dimension for the response register above which $p_n(r)$ is constant?
Perhaps $n+1$?  What happens to $p_{2j-1}(r)$ as $j\to\infty$ for 
fixed $r$?  Does it decrease to $1/2$?  Or to something larger?  What
constitutes a good, or optimal, choice of permutations and initial
response register state?

\medskip
\noindent{\bf Acknowledgements}

%--------|---------|---------|---------|---------|---------|---------|
This work has been partially supported by the National Science 
Foundation under grant ECS-0202087 and by the Defense Advanced 
Research Projects Agency as part of the Quantum Entanglement Science
and Technology program under grant N66001-09-1-2025.

\medskip
\global\setbox1=\hbox{[00]\enspace}
\parindent=\wd1

\noindent{\bf References}
\bigskip

\parskip=0pt
\item{[\HunzikerMeyer]}
%--------|---------|---------|---------|---------|---------|---------|
M. Hunziker and \dajm,
``Quantum algorithms for highly structured search problems'',
\QIP\ {\bf 1} (2002) 145--154.

\item{[\Angluinsurvey]}
%--------|---------|---------|---------|---------|---------|---------|
See, \eg,
D. Angluin,
``Computational learning theory:  Survey and selected bibliography'',
in
{\sl Proceedings of the Twenty-Fourth Annual ACM Symposium on Theory
     of Computing\/}
(New York:  ACM 1992) 351--369.

\item{[\AngluinSlonim]}
%--------|---------|---------|---------|---------|---------|---------|
D. Angluin and D. K. Slonim, 
``Randomly fallible teachers:  learning monotone DNF with an 
  incomplete membership oracle'', 
\ML\ {\bf 14} (1994) 7--26.

\item{[\SloanTuran]}
%--------|---------|---------|---------|---------|---------|---------|
R. H. Sloan and G. Tur\'an, 
``Learning with queries but incomplete information'', 
in 
{\sl Proceedings of the Seventh Annual ACM Workshop on Computational 
     Learning Theory\/}
(New York:  ACM 1994) 237--245.

\item{[\Shor]}
%--------|---------|---------|---------|---------|---------|---------|
\shor,
``Algorithms for quantum computation:  discrete logarithms and 
  factoring'',
in S. Goldwasser, ed.,
{\sl Proceedings of the 35th Symposium on Foundations of Computer 
Science}, Santa Fe, NM, 20--22 November 1994
(Los Alamitos, CA:  IEEE Computer Society Press 1994) 124--134;\hfb
\shor,
``Polynomial-time algorithms for prime factorization and discrete 
  logarithms on a quantum computer'',
{\tt quant-ph/9508027};
\SIAMJC\ {\bf 26} (1997) 1484--1509.

\item{[\HSPsurvey]}
%--------|---------|---------|---------|---------|---------|---------|
C. Moore, D. Rockmore, A. Russell and L. J. Schulman, 
``The power of strong Fourier sampling:  quantum algorithms for affine
  groups and hidden shifts'',
{\tt quant-ph/0503095};
\SIAMJC\ {\bf 37} (2007) 938--958.

\item{[\Grover]}
%--------|---------|---------|---------|---------|---------|---------|
\grover,
``A fast quantum mechanical algorithm for database search'',
in {\sl Proceedings of the Twenty-Eighth Annual ACM Symposium on 
  the Theory of Computing},
Philadelphia, PA, 22--24 May 1996 
(New York:  ACM 1996) 212--219;\hfb
\grover, 
``Quantum mechanics helps in searching for a needle in a haystack'', 
{\tt quant-ph/9706033};
\PRL\ {\bf 79} (1997) 325--328.

\item{[\HMPPR]}
%--------|---------|---------|---------|---------|---------|---------|
M. Hunziker, \dajm, J. Park, J. Pommersheim and M. Rothstein,
``The geometry of quantum learning'',
{\tt quant-ph/0309059};
\QIP, DOI 10.1007/s11128-009-0129-6 (online 23 September 2009).

\item{[\BernsteinVazirani]}
%--------|---------|---------|---------|---------|---------|---------|
\bv,
``Quantum complexity theory'',
in {\sl Proceedings of the 25th ACM Symposium on Theory of Computing},
San Diego, CA, 16--18 May 1993
(New York:  ACM Press 1993) 11--20;\hfb
\bv,
``Quantum complexity theory'',
\SIAMJC\ {\bf 26} (1997) 1411--1473.

\item{[\BargZhou]}
%--------|---------|---------|---------|---------|---------|---------|
A. Barg and S. Zhou,
``A quantum decoding algorithm for the simplex code'', 
in
{\sl Proceedings of the 36th Annual Allerton Conference on
     Communication, Control and Computing}, 
Monticello, IL, 23--25 September 1998 (UIUC 1998) 359--365.

\item{[\Helstrom]}
%--------|---------|---------|---------|---------|---------|---------|
C. W. Helstrom,
``Detection theory and quantum mechanics'',
\IC\ {\bf 10} (1967) 254--291.

\item{[\YKL]}
%--------|---------|---------|---------|---------|---------|---------|
H. P. Yuen, R. S. Kennedy and M. Lax,
``On optimal quantum receivers for digital signal detection'',
\PIEEEL\ {\bf 58} (1970) 1770--1773.

\item{[\Kennedy]}
%--------|---------|---------|---------|---------|---------|---------|
R. S. Kennedy,
``On the optimal receiver for the $M$-ary pure state problem'',
{\sl MIT Res.\ Lab.\ Electron.\ Quart.\ Prog.\ Rep.}\ {\bf 110}
(15 July 1973) 142--146.

\item{[\Helstrombook]}
%--------|---------|---------|---------|---------|---------|---------|
C. W. Helstrom,
{\sl Quantum Detection and Estimation Theory\/}
(New York:  Academic 1976).

\item{[\Plotkin]}
%--------|---------|---------|---------|---------|---------|---------|
M. Plotkin,
``Binary codes with specified minimum distance'',
\IRETIT\ {\bf 6} (1960) 445--450.

\item{[\Deutsch]}
%--------|---------|---------|---------|---------|---------|---------|
D. Deutsch,
``Quantum theory, the Church-Turing principle and the universal 
  quantum computer'',
\PRSLA\ {\bf 400} (1985) 97--117.

\item{[\DeutschJozsa]}
%--------|---------|---------|---------|---------|---------|---------|
D. Deutsch and R. Jozsa,
``Rapid solution of problems by quantum computation'',
\PRSLA\ {\bf 439} (1992) 553--558.

\item{[\Jozsa]}
%--------|---------|---------|---------|---------|---------|---------|
R. Jozsa,
``Quantum factoring, discrete logarithms, and the hidden subgroup 
  problem'',
{\tt quant-ph/0012084};
{\sl Computing Science Eng.}\ {\bf 3} (2001) 34--43.

\item{[\CEMM]}
%--------|---------|---------|---------|---------|---------|---------|
R. Cleve, A. Ekert, C. Macchiavello and M. Mosca,
``Quantum algorithms revisited'',
{\tt quant-ph/9708016};
\PRSLA\ {\bf 454} (1998) 339--354.

\item{[\Cauchy]}
%--------|---------|---------|---------|---------|---------|---------|
A. L. Cauchy, 
{\it Exercises d'analyse et de physique math\'ematique}, 
Tome {\bf 3} (Paris 1844) 151--252.

\item{[\SKIH]}
%--------|---------|---------|---------|---------|---------|---------|
M. Sasaki, K. Kato, M. Izutsu and O. Hirota,
``Quantum channels showing superadditivity in classical capacity'',
{\tt quant-ph/9801012};
\PRA\ {\bf 58} (1998) 146--158.

\bye